\documentstyle[preprint,aps,epsfig, multicol]{revtex}
\tighten
\draft
\widetext
\input{epsf.sty}
\newcommand{\nc}{\newcommand}
\nc{\beq}{\begin{equation}}
\nc{\eeq}{\end{equation}}
\nc{\beqa}{\begin{eqnarray}}
\nc{\eeqa}{\end{eqnarray}}
\nc{\ba}{\begin{eqnarray}}
\nc{\ea}{\end{eqnarray}}
\nc{\r}{\rho}
\nc{\rb}{\bar{\rho}}
\nc{\p}{\phi}
\nc{\pb}{\bar{\phi}}
\nc{\cb}{\bar{c}}
\nc{\s}{\sigma}
\nc{\k}{\kappa}
\nc{\Sb}{\bar{S}}
\nc{\kb}{\bar{\kappa}}

\begin{document}

\title{Closer towards inflation in string theory}
\vspace{1cm}

\author{Hassan Firouzjahi \footnote{Electronic mail:
firouzh@mail.lepp.cornell.edu}
and S.-H.~Henry Tye\footnote{Electronic mail:
tye@mail.lepp.cornell.edu}}
\vspace{0.5cm}
\address{Laboratory for Elementary Particle Physics, Cornell University, 
Ithaca, NY 14853}

\date{December 1, 2003}

\bigskip

\medskip
\maketitle

\begin{abstract}

In brane inflation, the relative brane position in the bulk of
a brane world is the inflaton.
For branes moving in a compact manifold, the approximate 
translational (or shift) symmetry is necessary to suppress the 
inflaton mass, which then allows a slow-roll phase for 
enough inflation. Following recent works, we discuss how inflation 
may be achieved in superstring theory. Imposing the shift symmetry,
we obtain the condition on the 
superpotential needed for inflation and suggest how this condition 
may be naturally satisfied.

\vspace{0.3cm}

Keywords : Inflation, Superstring Theory, Cosmology
\end{abstract}

\section{Introduction}

It is very likely that our universe has gone through an 
inflationary epoch. In slow-roll inflation 
\cite{Guth:1980zm}
the slow-roll condition 
imposes a very strong constraint on inflationary model 
building; in particular, it requires a very small inflaton
mass. In simple brane world scenarios inspired by string 
theory, brane inflation is a natural way to achieve this 
\cite{Dvali:1998pa}. For a BPS brane moving in the bulk
of the brane world, the translational, or shift, symmetry implies a 
massless inflaton, that is, the brane position is a massless 
Goldstone boson.
A weak brane-brane interaction breaks this symmetry slightly,
so that the shift symmetry is still an approximate symmetry, 
or equivalently, the inflaton is a pseudo-Golstone boson. 
This results in a relatively flat inflaton potential, allowing 
enough inflation to take place before the branes collide and 
give birth to the radiation-dominated big bang. 

It is obviously important to see how brane inflation may be 
realized in a more realistic situation, 
where the compactification of the extra 
dimensions in superstring theory is dynamically stabilized. 
Recently, there was just such an attempt (KKLMMT) \cite{Kachru:2003sx}. 
In the supergravity approximation, 
KKLMMT find that the K\"{a}hler potential contribution to the 
inflaton mass is of order $m_{\phi}^2 \simeq 2H^2/3$, while the 
slow-roll condition requires that $|m_{\phi}^2| \le  H^2/100$.
The slow-roll conditon may be reached if the superpotential
contribution to $m_{\phi}^2$ cancels that from the K\"{a}hler potential.
In this note, we keep track of the shift symmetry,
and identify the condition imposed by the shift symmetry on
$W$ and the inflaton potential $V$. We then suggest that the shift 
symmetry in $W$ required for slow-roll inflation may appear
quite naturally. In short, symmetry argument alone may be enough to 
tell us whether
slow-roll inflation will take place or not, even though an 
explicit determintion of $W$ and $V$ may be very difficult.

Here is a brief review and a summary of the scenario. 
Start with a 4-fold Calabi-Yau manifold in F-thoery,
or, equivalently, a type IIB orientifold compactified on a 
3-fold Calabi-Yau manifold 
with fluxes to stabilize all but the volume 
modulus \cite{Giddings:2001yu}. For large volume, supergravity
provides a good description.
Next, one introduces a non-perturbative
superpotential $W$ that stabilizes the volume modulus in a
supersymmetric AdS vacuum. The introduction of $\bar D$3-branes 
in a warped type IIB background
breaks supersymmetry and lifts the AdS vacuum to a metastable 
de Sitter (dS) vacuum \cite{Kachru:2003aw} (the KKLT vacuum). 
To realize inflation, KKLMMT introduces a $D$3-$\bar D$3-brane 
pair, whose vacuum energy drives inflation
\cite{Burgess:2001fx}.
The $\bar D$3-brane is fixed with the other $\bar D$3-branes 
and the inflaton $\phi$ (complex $\phi_i$, $i=$1,2,3) is 
the (6-dimensional) position (relative to the $\bar D$3-branes)
of the $D$3-brane.
Consider the potential for the mobile $D$3-brane in the
$D$3-${\bar D}$3-brane inflationary scenario 
\ba
\label{fullV}
V(\rho,\phi) &=& V_F(K(r),W(\rho,\phi)) + \frac{B}{r^2} + 
V_{D{\bar D}}(\phi)\\
K(r)&=& -3 \ln(2r)= -3 \ln (\rho +{\bar \rho} -\phi {\bar \phi})
\ea
where $r$ is the physical size of the compactified volume and
$\rho$ the corresponding bulk modulus. 
The form of the K\"{a}hler potential came from Ref\cite{DeWolfe:2002nn},
where $r$ is invariant under the constant shift symmetry:
\ba
\label{shift}
\phi_i \rightarrow  \phi_i + c_i, \quad  
\r \rightarrow  \r + {\bar c}_i \phi_i +  \sum c_i{\bar c_i}/2 \ . 
\ea
$V_F(K(r),W(\rho,\phi))$ is the F-term potential, where the
superpotential $W$ is expected to stabilize the volume modulus
in an AdS supersymmetric vacuum. KKLT then introduces $\bar D$3-branes
(the $Br^{-2}$ term) in a Klebanov-Strassler throat that breaks
supersymmetry to lift the AdS vacuum to a de Sitter vacuum (a
metastable vacuum with a very small cosmological constant and a
lifetime larger than the age of the universe). Note that this term 
leaves the shift symmetry intact.
The $D$3-$\bar D$3 potential
$V_{D{\bar D}}(\phi)$ is presumably very weak due to warped geometry.
This inflaton potential is designed to break the shift symmetry 
slightly, so inflation can end after slow-roll.

To realize slow-roll inflation in the early universe, $V(\rho,\phi)$
must be very flat in some
of the $\phi$ directions around its minimum where the
compactification size is stabilized, that is, some of the
$\phi$ components must have very small masses $|m_{\phi}^2| \le  H^2/100$,
where $H$ is the Hubble constant during the inflationary epoch, while
the remaining components can be more massive.
Hopefully inflation takes place as the $D$3-brane moves slowly towards
the $\bar D$3-branes.
Inflation ends when they collide and annihilate.
In this scenario, the inflaton potential comes from 2 sources :
(1) the inter-brane potential $V_{D{\bar D}}(\phi)$, which is 
sufficiently weak due to the warped geometry, and 
(2) the $D$3-brane coupling inside $V_F$.
KKLMMT showed that $K$ inside $V_F$ alone contributes $2H^2/3$ to 
$m_{\phi}^2$ for all components of $\phi$, 
while the non-perturbative $W(\rho)$ they used to stabilize $r$, as
it does not depend on $\phi$, does not contribute to $m_{\phi}^2$.
Since the contributions to $m_{\phi}^2$ by the remaining terms in
$V(\rho,\phi)$ around its minimum is negligible,
$m_{\phi}^2 \simeq 2H^2/3$, so the slow-roll condition 
$\eta \simeq m_{\phi}^2/H^2 \le 1/60$ is not satisfied.
Without a better knowledge of $W$, this requires a 1 part 
in 100 fine-tuning. Actually, small $\eta$ is required during 
the whole inflationary epoch as the
$D$3-brane moves slowly towards the $\bar D$3-branes, not just 
at a single value of $\phi$. This 
condition is much more stringent than the 1 in 100 fine-tuning.
Such a condition must come from a symmetry argument.
Here we keep track of the shift symmetry needed for inflation
and identify the condition imposed by the shift symmetry on
$W$.

To realize slow-roll inflation, we should consider a modified $W$ in
a way that some of the shift symmetry (\ref{shift})
is left intact in $V_F$.
Rcall that $W(\rho)$ used in KKLT is obtained from non-perturbative 
effects in the absence of $\phi$.
In the presence of a mobile $D$3-brane, some $\phi$ dependence
in $W$ is quite natural. Motivated by the
shift symmetry, we consider
\beq
\label{Wnsp}
W(\rho,\phi) = - w_0 + A e^{-a(\rho - \kappa_i \phi_i^2/2)}
\eeq
where $w_0, A$ and $a$ are real parameters. In the absence of $\phi$, 
Eq(\ref{Wnsp}) reduces to the $W(\rho)$ used in KKLT. 
For generic parameters $\kappa_i$, it is clear that
slow-roll conditions is not satisfied. However, demanding that
the volume modulus is stabilized in a supersymmetric vaccuum in $V_F$,
we show that either $\kappa_i=0$, which yields the original KKLT vacuum
(which has no shift symmetry left), or 
\beq
SUSY \rightarrow |\kappa_i|=1 \rightarrow \eta \simeq 0
\eeq
that is, the slow-roll condition is satisfied. More precisely,
the supersymmetric vacuum has a degeneracy where
3 components of $\phi$ are massless while the other 3 have
$m^2 \simeq 2H^2/3$, so, during inflation, the $D$3-brane will 
fall rapidly to the 
minimum in the massive $\phi$ directions and then move slowly 
along the flat directions. We may set $\kappa_i=1$ without
loss of generality.

More generally, the potential $V_F$ from a $W$ of 
the form (for any real constants $\beta_i$) 
\beq
W= F(\sigma)e^{i \beta_i \phi_i} \quad \quad \sigma=\rho-\phi^2/2
\eeq
has the remnant shift symmetry ($\phi_i \rightarrow \phi_i + Re (c_i)$)
that leaves the 3 real components of $\phi$ massless and so
slow-roll inflation can take place. This is the condition
on the $D$3-$\bar D$3-brane inflation in this scenario. 
(More precisely, we actually need only one of the $|\kappa_i|=1$.)

The question reduces to how natural/likely for $W$ to 
depend (up to the phase factor) 
only on $\sigma$. To address this question, let us consider 
the non-perturbative interactions of the $SU(N)$ super-Yang-Mills theory
due to the $N$ $D$7-branes wrapping a 4-cycle in the 
Calabi-Yau manifold. In the absence of the $D$3-brane,
$8 \pi^2/g_{YM}^2= Re (\rho)$. $SU(N)$ strong interaction
generically generates $W(\rho)=A e^{-a\rho}$ in the absence of $\phi$.
The presence of a $D$3-brane introduces open string modes 
stretching between the $D$3-brane and the $N$ $D$7-branes, 
which generically yields a bi-fundamental mode that behaves as a 
quark charged under 
$SU(N)$. The mass of the quark is $m=|\phi_{(D7)}|$, where $\phi_{(D7)}$
is $\phi$ measured with respect to the position of the wrapped $D$7-branes.
(See Figure 1.) 
So $W$ should depend on $\phi$ as well. There should be a subspace
of $\phi$ which has the same quark mass $m$ and so the same value 
for the $W(\rho,\phi)=W(\Lambda,m)$, where $\Lambda$ is the scale 
for $SU(N)$. Naively, this subspace might be the $S^5$ 
surface around the $D$7-branes with a radius given by $m=|\phi_{(D7)}|$,
that is, the components of $\phi$ tangent to this surface are 
massless Goldstone modes.
However, as pointed out above, supersymmetry and the holomorphicity 
of $W$ allows at most 3 flat directions. 
For slow-roll inflation, a single flat direction is enough. 
In this sense, the presence of a remnant shift symmetry in $W$ is 
quite natural. Once one shows that the shift symmetry is present 
in the supersymmetric vacuum, the slow-roll condition is generically
satisfied. The precise form of $W$ may not be necessary.

Recently, Ref.\cite{Hsu:2003cy} discusses the shift symmetry
for the $D$7-brane in the $D$3-$D$7 inflationary 
scenario \cite{Dasgupta:2002ew}. 
We shall comment on this interesting work.

\vspace{2cm}
\centerline{\epsfxsize=10cm\epsfbox{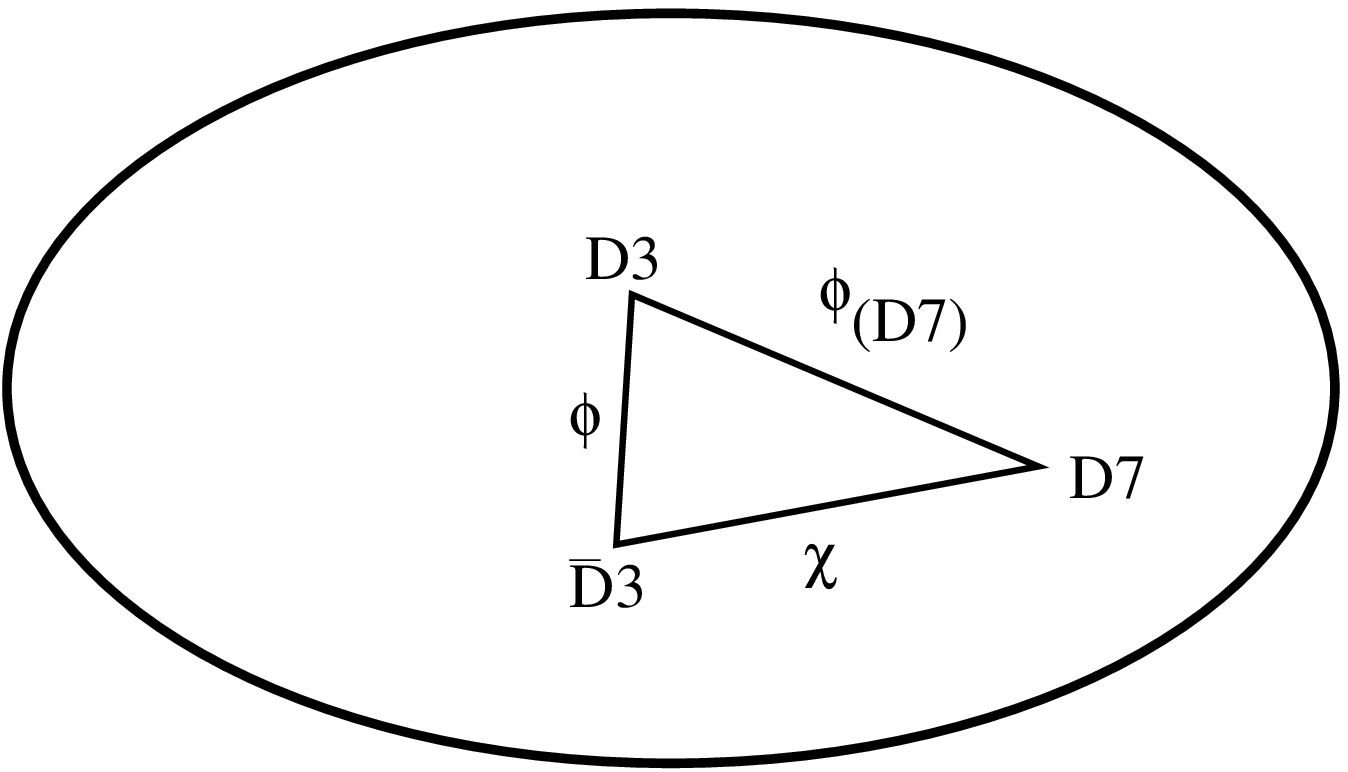}}
\vspace{0.1in}
\noindent {{\bf FIG}. 1. 
The relative positions of the wrapped $D$7-branes, the 
$D$3-brane and the ${\bar D}$3-branes inside a Calabi-Yau manifold. $\p$ is 
the position of the $D$3-brane relative to the ${\bar D}$3-brane.}
\label{1}

\vspace{1cm}

\section{The Shift Symmetry and $W$}

Given the K\"{a}hler potential and the superpotential $W$, the 
potential $V_F$ is given by
\beq
V_F=e^K \left( K^{i{\bar j}} D_iW\overline{D_ jW} -3 W \overline{W} \right)
\eeq
where 
\ba
K_i = \frac{\partial K}{\partial \varphi_i},    \quad
D_iW = \frac{\partial W}{\partial \varphi_i} + K_iW 
\ea
Physics is determined by the particular combination $G$,
\ba
\label{G}
G=K+ \ln(W) +\overline{\ln(W)} 
\ea
since it is invariant under the K\"{a}hler transformation
\ba
K \rightarrow K+ \xi(\varphi_i)+ \overline{\xi(\varphi_i)}, 
\quad W \rightarrow W e^{-\xi(\varphi_i)} \ .
\ea
The potential is given uniquely in term of $G$
\ba
\label{V}
 V_F =e^{G}\left[G^{i\bar{j}}~\frac{\partial G}{\partial \varphi_i}~ 
\overline{\frac{\partial G}{\partial \varphi_j}} - 3 \right]
\ea
where $G_{i \bar{j}}=K_{i \bar{j}}$.

For a $D$3-brane with position $\phi$ in a compact manifold, 
the K\"{a}hler potential is \cite{DeWolfe:2002nn}
\beq
\label{Kahler}
     K=-3~ \ln (\r+\rb-k(\p,\pb))
\eeq
where $\rho$ is the bulk modulus and $\phi$ is the 
position of a mobile $D$3-brane (in 6-dimensions, we have
$\phi_i$ ($i=$1,2,3)).
The physical radius of the Calabi-Yau manifold is given by
\beq
    2r ~ = \r +\rb - k(\p,\pb) \ .
\eeq
Note that $r$ is invariant under the transformation :
\beq
\r \rightarrow \r + f(\phi), \quad \rb \rightarrow \rb + {\bar f(\phi)},
\quad k(\p,\pb) \rightarrow k(\p,\pb) + f + {\bar f} \ .
\eeq
Let us focus on the simpliest choice of $k(\p,\pb)$, 
\beq
k(\p,\pb)=\p \pb = \sum \p_i\pb_i \ .
\eeq
So the above symmetry reduces to the shift symmetry (\ref{shift}).
(Compared to terms with higher powers of $\p \pb$ in $k(\p,\pb)$,
the above leading $\p \pb$ term contributes dominantly to $\eta$.)
The supergravity effective action is valid only for large $r$.
This is the regime we are interested in. 
For a constant $W$, it is easy to check that $V_F=0$. This is an example of 
no-scale SUGRA and we have the full shift symmetry. 
However, we are interested in supersymmetric vacua. 
Imposing the supersymmetry condition, 
\ba
\label{supersymmetry}
    D_{i}W&=&\frac{\partial W}{\partial \p_i} ~ + W \frac{\partial K}
{\partial \p_i} ~ =0 \nonumber \\
D_{\r}W&=&0 \nonumber
\ea
we obtain
\ba
\frac{\partial}{\partial \p_i} \ln(W)  &=&  
\frac{-3\pb_i}{\r +\rb -\p\pb} \\\nonumber
\frac{\partial}{\partial \r} \ln(W)  &=& \frac{3}{\r +\rb -\p\pb}  \ .
\ea
Let us first consider a toy model with maximum shift symmetry and 
where the bulk modulus is not stabilized. 
In this case we can treat Eq.(\ref{supersymmetry}) as differential 
equations.
Imposing the holomorphic condition on the supersymmetric $W_s$, we obtain 
\beq
\label{W}
      W_s ~ = w_0 ~ ( \r - \kappa_i\p_i^2/2)^{\frac{3}{2}}
\eeq
where the constant $w_0$ is the normalization and $|\kappa_i|=$1.
The line of the continues solutions of Eq(\ref{supersymmetry}) is given 
by $\k_i=\pb_i/\p_i$, $\r=\rb$. 
This leaves a subset of the shift symmerty intact: the 
shift symmetry (\ref{shift}) where the $c_i$ satisfies
$\kappa_i= {\bar c}_i/c_i$.  
Without loss of generality, we can choose $\kappa_i=$1 by the field 
redefinition $\p_{i~new}=\sqrt{\k_i}\p_i$. At the supersymmetric 
AdS vacuum, $V_F=-\frac{3}{8} w_0^2 $ and
the remnant shift symmetry is given by real $c_i$~ $(c_i=\bar{c_i})$ :
\beqa
\label{shift phi}
    \p_i &\rightarrow & \p_i + c_i \nonumber\\
    \pb_i &\rightarrow & \pb_i + c_i \nonumber \\
    \r &\rightarrow & \r + \p c + c^2/2 \ .
\eeqa
Eq.(\ref{W}) suggests that instead of ($\rho$,$\phi_i$), it is 
natural to use ($\sigma$,$\phi_i$), where 
\beq
\sigma= \r- \frac{\phi^2}{2} = \r - \sum \frac{\phi_i^2}{2} \ .
\eeq
That is, $\sigma$ is invariant under the above Re($c_i$) shift.
The transition from $(\r,\p)$ to 
$(\s,\p)$ is always nonsingular. 
In term of ($\sigma$,$\phi$), 
$W_s(\sigma) = \omega_0 \sigma^{\frac{3}{2}}$ and
\beq
2r= \sigma + {\bar \sigma} +\frac{(\phi-{\bar \phi})^2}{2} \ .
\eeq
We see that the remnant $c=\bar c$ shift symmetry will be left intact 
for any $W(\sigma)$. 

Next, we like to find the most general $W$ that stabilizes the 
compactification volume in a supersymmetric vacuum with some 
remnant shift symmetry.
The physics is more transparent in the $(\sigma,\p)$ coordinate. 
Let $\ln W (\s,\p)= f(\s,\p)$, the solution at the supersymmetric 
minimum is given by
\ba
\label{fp} 
f_i|_0\equiv f_{\p_i}|_0&=&\frac{3 (\p_i - \pb_i )|_0 }{2r_0} \\\nonumber 
f_\s|_0&=&\frac{3}{2r_0}
\ea
where the subscript $0$ indicates the 
supersymmetric minimum point.
Under the Re$(c_i)$ shift transformation, these equations remain invariant 
only if
\ba
\label{fs} 
f_{\s i}=f_{ij}=0 \ .
\ea
So a solution which has a degenerate supersymmetric minimum
has the form
\ba
f(\s,\p)=h(\s) + i \beta_i \p_i 
\ea
where $h(\s)$ is any function of $\s$, except being a pure constant, 
as can be 
seen from Eq.(\ref{fp}). Furthermore, Eq.(\ref{fp}) indicates that 
$\beta_i$ are real numbers. 
Finally, the form of the superpotential is (for real $\beta_i$)
\ba
\label{final}
W(\sigma,\phi_i)=F(\sigma)~e^{i\beta_i \p_i} \ .
\ea
The above superpotential is not Re$(c_i)$ shift invariant, because 
of the phase term. However, the combination 
$\ln W + \overline{\ln W}$ is. This is exactly the contribution 
of the superpotential to $G$, as can be seen from Eq.(\ref{G}).
That is, the minimum (the AdS supersymmetric vacuum) of $V_F$ is 
Re$(c_i)$ shift invariant.

As a comment, let us look at the kinetic part of the action in 
terms of $(\sigma,\p)$. Calculating 
the K\"{a}hler metrics $K_{i \bar{j}}$ from the K\"{a}hler potential 
(\ref{Kahler}), we find
\ba
\label{kinetic}
     {\cal L}_{k}&=&\frac{1}{2}\sqrt{g} 
     K_{i \bar{j}}\partial_{\mu}\varphi_{i}{\partial^{\mu}
\bar \varphi_{j}}      \\\nonumber 
     &=&\frac{3 \sqrt{g}}{8r^2} \left(\partial_{\mu}\r
\partial^{\mu}\rb+(2r\delta_{ij}+\pb_i\p_j)\partial_{\mu}\p_i
\partial^{\mu}\pb_j -\p_i\partial_{\mu}\r\partial^{\mu}\pb_i
-\pb_i\partial_{\mu}\p_i\partial^{\mu}\rb \right) \\\nonumber
&=&\frac{3\sqrt{g}}{8r^2} ([2r \delta_{ij}-(\p_i-\pb_i)(\p_j-\pb_j)]
\partial_{\mu}\p_i \partial^{\mu}\pb_j
+(\p_i-\pb_i) (\partial_{\mu}\p_i \partial^{\mu}\bar{\sigma}
 -\partial_{\mu}\sigma \partial^{\mu}\pb_i)
+\partial_{\mu}\sigma \partial^{\mu}\bar{\sigma}) .
\ea
By construction $\sigma$ is Re$(c_i)$ shift invariant. 
It is clear that the kinetic term ${\cal L}_{k}$ 
is shift invariant, when expressed in
either ($\r,\p$) or $(\sigma,\phi)$ basis. In the $(\sigma,\phi)$
basis, each term in ${\cal L}_{k}$ is explicitly shift invaraiant.

\vspace{0.3cm}

\section{K\"{a}hler modulus stabilization and dS vacuum}

Following KKLT, we shall introduce non-perturbative quantum 
corrections to the superpotential
that can lead to supersymmetric AdS vacua in which the K\"{a}hler
modulus is fixed. Such a correction can come from a number of 
sources : Euclidean $D$3-branes wrapping some 4-cycles 
\cite{Witten:1996bn}, or strongly coupled $SU(N)$ super-Yang-Mills theory 
associated with the $N$ $D$7-branes wrapping a 4-cycle
in a Calabi-Yau manifold in a Type IIB oreintifold (or F theory).
The gauge coupling is related to the dilaton and the size of 
the 4-cycle that the $D$7-branes wrap on. Since the dilaton 
and the complex moduli are already stabilized \cite{Giddings:2001yu},
the gauge coupling depends only on the size of the 4-cycle, which 
is proportional to the size of the Calabi-Yau manifold.
So $8 \pi^2/g_{YM}^2= Re (\rho)$. $SU(N)$ QCD
generically generates $W(\Lambda)=W(\rho)$, where 
$\Lambda$ is the QCD scale of the strongly interacting
$SU(N)$ super-Yang-Mills theory.
In the absence of $\phi$ and in the large volume regime, 
$W(\rho)=A\exp(-a\rho)$. For large $N$, $a \simeq 1/N$.

The presence of a $D$3-brane introduces $\phi$, its position.
As shown in Figure 1, let $\chi$ be the separation between 
the $D$7-branes and the ${\bar D}$3-branes. Let $\phi_{(D7)}$
be the position of the $D$3-brane measured relative to
the $D$7-branes, so $|\phi_{(D7)}|$ is the separation distance 
between the $D$3-brane and the $D$7-branes. Then
$\phi=\phi_{{\bar D}3}$ is the inflaton. 
(Strictly speaking, the ${\bar D}$3-branes are sitting at 
$\varphi_0$, instead of $\phi=$0. When the $D$3-brane reaches 
$\phi\simeq\varphi_0$, tachyons appear and inflation ends rapidly, 
as in hybrid inflation. Since generically $|\phi |>>\varphi_0$,
this distinction is not relevant for our discussion here.) 
Using the shift symmetry (\ref{shift}), we see that
$\rho=\rho_{{\bar D}3}$ and $\rho_{(D7)}$ are different but
related:
\ba 
\phi_i &=& \phi_{(D7) i} + \chi_i \\\nonumber
\rho &=& \rho_{(D7)} + {\bar \chi} \phi_{(D7)} + 
\frac{\chi {\bar \chi}}{2} \ .
\ea
so that the physical size of the Calabi-Yau manifold, 
$2r ~ = \r +\rb - \p\pb = \rho_{(D7)} + {\bar \rho}_{(D7)}
-\phi_{(D7)}{\bar \phi}_{(D7)}$, is invariant under the change in 
coordinates, as it should be.

The presence of a $D$3-brane introduces open string modes 
stretching between the $D$3-brane and the $N$ $D$7-branes, which 
generically yields a bi-fundamental mode that behaves 
as a quark charged under $SU(N)$. The mass of the quark is 
\beq
\label{mquark}
m=|\phi_{(D7)}|
\eeq
which may be suitably modified by the warped geometry.
Since the strong interaction dynamics of the super-Yang-Mills theory
depends on the quark mass $m$, $W$ must depend on 
$\phi$ as well \cite{Ganor:1996pe}. There should be a subspace
of $\phi$ which has the same quark mass $m$ and so the 
same value for the $W(m,\Lambda)=W(\rho,\phi)$; that is, the 
non-perturbative dynamics depends on $\phi$ only through 
Eq.(\ref{mquark}). This is the degeneracy we are interested in. 
Treating (naively) the 4-cycle as a point in the Calabi-Yau manifold,
one may expect this subspace to be the $S^5$ surface with 
radius given by $m=|\phi_{(D7)}|$. That is, the 5 modes tangent 
to this surface are massless Goldstone modes. 
Depending on how the 4-cycle is actually embedded in the 
Calabi-Yau manifold, it is likely that some of the Goldstone 
modes are lifted. As we have shown, supersymmetry and the 
holomorphicity of $W$ allows at most 3 flat directions in the 
supersymmetric vacuum. 
Still, the above argument suggests that the flat 
directions should be orthogonal to the $\phi_{(D7)}$ direction.
Although the interaction between $D$3-$\bar D$3-branes are weak
(and given by $V_{D \bar D}$), the interaction between $\bar D$3-branes and
$D$7-branes is important for the $SU(N)$ strong interaction.
There are also  bi-fundamental modes stretching between the $\bar
D$3-branes and the $D$7-branes that behave like additional
quarks of $SU(N)$, with mass $m'$. Presumably, the stabilization of the
dilaton and complex structure moduli fixes $m'$, up to an overall volume
scaling which is a function of $\rho$. This means that $W(m,m',\Lambda)=
W(\rho, \phi)$.
In general, $W$ can be quite complicated. However, the above argument
suggests that its supersymmetric vacuum should have a number of 
flat directions. 
For small $\phi$, this suggests that the flat directions should be
orthogonal to the $\chi$ direction. 
A simple extension suggests (with real positive parameters $w_0$,
$A$ and $a$)
\beq
\label{Wsnp}
W= -w_0+A e^{-a \rho} \rightarrow 
 \left( -w_0+Ae^{-a\sigma} \right)~e^{i \sqrt{\kappa_i}\beta_i \phi_i}
\eeq
that is, the supersymmetric non-perturbative correction to the 
superpotential $W$ is a funcion of 
\ba
\sigma &=& \rho - \kappa_i \phi_i^2/2 \\\nonumber
\kappa_i &=& - {\bar \chi}_i/\chi_i \\\nonumber
\beta _i &=& {\bar \beta}_i \ .
\ea
Other possible solutions lift at least one of the flat directions:
for example, we may end up with $\kappa_1=0$, $|\kappa_2|=|\kappa_3|=1$
or $\kappa_1=\kappa_2=0$, $|\kappa_3|=1$. For slow-roll inflation, 
all we need is one flat direction. On the other hand, the original 
solution of KKLT, with $\kappa_1=\kappa_2=\kappa_3=0$, has no 
flat direction. 

Let us consider the above solution with 3 flat directions.
In the absence of $\phi$ (say after inflation and the annihilation 
of the $D$3-$\bar D$3-brane pair), the form of the potential reduces
to the well-known one.
In the coordinate where $\chi_i$ are purely imaginary,
$\sigma = \rho - \phi^2/2$.
In the large volume limit and small $\phi$, the difference 
between $\rho$ and $\sigma$ is 
only a small quantitative correction. 
Once the dynamics are better 
understood, one may be able to put more precise constraints on 
the 4-cycles that $D$7-branes can wrap. It will be most interesting 
to see if the inclusion of a mobile $D$3-brane yields the above form 
for $W$. Here, let us simply assume that there are 
such supersymmetric non-perturbative correction that leaves the 
shift symmetry intact. Now we have
\ba
G=\ln \left(\frac{|(-\omega_0+ A e^{-a \sigma})e^{i \beta \phi}|^2}
{\left(\sigma +\bar{\sigma}+\frac{(\p -{\bar \phi})^2}{2}\right)^3} \right) \ .
\ea
The supersymmetric minima is given by $G_{\s}=G_{\p}=0$,
which for real $\s$ has the solution 
\ba
Im(\p_{i0}) &=& {r_0}\beta_i/3 \nonumber\\
\omega_0 &=&(1+\frac{2a}{3}r_0 )Ae^{-a(r_0+\frac{\beta^2}{9}r_0^2)} \\\nonumber
V_{AdS} &=&-\frac{1}{6r_0}a^2A^2e^{-2a r_0}e^{-\frac{2}{3}\beta^2 r_0
(1+\frac{a}{3}r_0)}
\ea

Next, to lift the AdS vacuum to a dS vacuum, we must
break supersymmetry. Again following KKLT, we introduce a 
$\bar D$3-brane (or more than one), with its position fixed by 
the fluxes.
(One must adjust the 5-form flux so tadpole cancellation remains valid.)
This means no additional translational moduli are introduced. 
The net effect is simply the addition of an energy density of 
the form $V_{{\bar D}3} = B/r^2$.
Since $r$ is invariant under the remnant shift symmetry, no 
modification of this term is needed here. 
To simplify the result we may consider the case where $\beta_i=0$, 
corresponding to $(\p_i-\pb_i)_0=0$. 
For Im($\p_i$)=0, the potential takes the form :
\ba
V&=&V_F(K,W)  + V_{{\bar D}3}\nonumber\\
&=&\frac{a A e^{-ar}}{2r^2}(-\omega_0+Ae^{-ar}+\frac{1}{3}a A re^{-ar}) 
+\frac{B}{r^2} \ .
\ea 
which is clearly independent of Re ($\phi_i$).
Of course, one must fine-tune to obtain a very small 
cosmological constant. One may choose another mechanism to 
lift the AdS vacuum to the dS vacuum. The shift symmetry is 
maintained as long as the lifting is a function of $r$ only.

\section{$D$3-${\bar D}$3 Inflation}

We can now introduce an extra $D$3-${\bar D}$3-brane pair, where the 
${\bar D}$3-brane is stuck with other ${\bar D}$3-branes while the
$D$3-brane moves towards them. The above analysis for the mobile 
$D$3-brane can be used for this additional $D$3-brane.
The resulting vacuum energy 
drives inflation. The remnant shift symmetry is broken only by the 
very weak $D$3-${\bar D}$3-brane interaction, namely, 
$V_{D{\bar D}}(\phi)$ in Eq.(\ref{fullV}).

In the case where $\beta_i=0$, the $D$3-brane rapidly falls to 
Im ($\phi$)$=0$ at the early stage of the inflationary epoch, 
and then moves slowly towards $\phi=0$. For very small $|\phi|$,
a tachyonic mode appears and inflation ends quickly, as in 
hybrid inflation. 
 
For the case where $\beta_i \ne 0$, the situation may be different,
since the ${\bar D}$3-brane is not sitting along the flat direction 
the $D$3-brane is moving.
For large $\beta_ir_0$, the $D3$ may not collide with the $\bar D3$, 
at least classically.
Let us consider the case when the $D$3-brane along the flat direction 
is close to the ${\bar D}$3-brane, as described by
the following simple potential:
\beq
V=\frac{1}{2}m_\p^2(\p-\frac{{r_0}\beta}{3} )^2-\frac{C}{\p^4}
+ constant
\eeq
where, to simplify the analysis, we consider only one component
of $\p$. The last term comes 
from the $V_{D{\bar D}}(\phi)$. Here $m_{\p} \sim H$.
In the absence of
$V_{D{\bar D}}(\phi)$, the $D3$-brane will move along $\p= Im(\p)
=r_0\beta/3$. However, $V_{D{\bar D}}(\phi)$ will tend to drive the 
$D3$-brane towards the $\bar D3$-brane. We like $V_{D{\bar D}}(\phi)$
to overcome $m_\p^2(\p-\frac{{r_0}\beta}{3} )^2/2$ so that
inflation can end quickly.

The condition that the $D3$-brane collides with $\bar D3$-brane
along its trajectory is that $V'> 0$, that is $V'$ has no local 
minimum away from $\p=0$. This in turn implies that 
\beq
{r_0}\beta < \frac{18}{5}(\frac{20~C}{m_\p^2})^\frac{1}{6} \ .
\eeq
To estimate this bound, we recall the KKLMMT scenario :
$C=\frac{4\pi^2}{{\cal N}^2}\varphi_0^8$, where ${\cal N}$ is 
the number 
of the $D3$ branes (or 5-form charge) in the original KKLT vacuum,
and $\varphi_0$ indicates the position of the $\bar D3$-brane in 
the Klebanov-Strassler throat. We require $\phi=\varphi_0$ for 
inflation to end.
We find 
\beq
{r_0}\beta < 4 \times 10^{-6}m_\p^{-1/3}.
\eeq
Assuming that inflation take places at the GUT scale and restoring the 
powers of $M_{Pl}$ in the calculations, we obtain 
${r_0} (\beta M_{Pl}) < 4~\times 10^4$, which is not vey restrictive.

\section{$D$3-$D$7} 

Ref\cite{Hsu:2003cy} considers the shift symmetry for the $D$7-branes. 
Using the K\"{a}hler potential for a $D$7-brane (described by $S$), 
they consider the D-term potential \cite{Burgess:2003ic} and the
shift symmetry in a supersymmetric vacuum.
The superpotential which has a supersymmetric minimum and is compatible 
with the shift symmetry $S \rightarrow S+c$ for real $c$ has the form 
\ba
W=e^{-\frac{1}{2}S^2}e^{i\gamma S}
\ea
where $S$ represents the position of the $D$7 brane on the 
Calabi-Yau manifold and 
$\gamma$ is a real number, similar to $\beta$ in D3 case.

For the system of non-interacting $D$3 and $D$7 brane, the form of the 
K\"{a}hler potential and the superpotential are expected to be
\ba
\label{D3D7s}
K&=&-3 \ln (\r +\rb-\p\pb)+\Sb S\nonumber\\
W&=& F(\s)e^{-\frac{1}{2}S^2}e^{i(\beta \p +\gamma S)} \ .
\ea
Now the system has four flat directions, three along the $(\p_i)$ 
subspace and one along Re$(S)$.


Following the above argument, we do expect that the $\phi$ in 
Eq.(\ref{D3D7s}) corresponds to $\phi_{(D7)}$ measured with respect 
to the initial $D$7-brane position. 
The inflaton for such a $D$3-$D$7 system is the relative positions 
of the $D$3-brane and the $D$7-branes. However, as discussed earlier,
we do expect a
non-trivial non-perturbative potential between the $D$3-brane and 
the $D$7-branes due to the quark mass in the 
super-Yang-Mills theory.

\section{Discussions}

Instead of Eq.(\ref{Wsnp}), one may also use the racetrack 
idea to stabilize the K\"{a}hler modulus
\cite{Escoda:2003fa}. In this scenario, there are more than one 
strongly interacting gauge groups. The number of flat directions 
may be reduced by the presence of multi-centers of super-Yang-Mills 
theories. It will be interesting to find the constraint on the 
super-Yang-Mills theories due to the shift symmetry.

There will be quark fields coming from the open string modes 
stretching between the ${\bar D}$3-branes and the $D$7-branes. 
They will also have impact on the non-perturbative dynamics of 
the super-Yang-Mills theory. The dynamical effect of this quark 
is similar,  
if not identical, to that for the quark field due to the $D$3-brane.
Since the open string modes between the $D$3-brane and the
${\bar D}$3-branes are not involved in the super-Yang-Mills dynamics,
one may expect that there is an almost flat direction that 
connects them, that is,
the dynamics may place the $D$3-brane along an almost flat direction 
that ends on the ${\bar D}$3-branes.
As we have shown, it is also possible that the ${\bar D}$3-branes
do not lie along the almost flat direction of the mobile $D$3-brane.
A little tuning may be needed to place the ${\bar D}$3-branes 
along the almost flat direction of the $D$3-brane. 

Suppose a fine-tuning is needed to obtain slow-roll inflation.
For the sake of discussion, let us say this is a 1 in 100 
fine-tuning. It is known that there are many metastable KKLT type
vacua, simply by changing the flux quanta. If the early universe
is composed of 100 (or more) causally disconnected regions
(a rather generic situation), each region may correspond to a 
different vacuum. If only one of them goes through a slow-roll
inflationary phase, this region grows by an 
exponential factor, compeletely dominates over the regions 
that do not inflate. So
it is exponentially unlikely that one will end up in anyone 
of the regions that do not inflate. 
That is, cosmologically, there is no fine-tuning.

To conclude, even though the determination of $W$ may be quite 
difficult, physics becomes more transparent when we view from 
the shift symmetry perspective, which alone may be enough to tell us 
whether slow-roll inflation will take place or not. If slow-roll brane 
inflation takes place, the inflationary properties may not be very 
sensitive to the details of $W$, but it does depend on the 
structure of the degeneracy of the supersymmetric vacuum.

\vspace{0.3cm}
 
We thank Cliff Burgess, Renata Kallosh, Louis Leblond, Shamit Kachru, 
Juan Maldacena,
Fernando Quevedo and Sandip Trivedi for valuable discussions.
This material is based upon work supported by the National Science
Foundation under Grant No.~PHY-0098631.

\end{document}